\documentclass[preprint,amsmath,amssymb,aps,showkeys,showpacs]{revtex4}
\usepackage[english]{babel}
\usepackage{graphicx}
\usepackage{graphics}
\usepackage{amsmath}
\usepackage{dcolumn}
\usepackage{amssymb}
\usepackage{bm}

\begin{document}
\title{Noninertial effects on nonrelativistic topological quantum scattering}
\author{H. F. Mota}
\email{hmota@fisica.ufpb.br}
\affiliation{Departamento de F\'isica, Universidade Federal da Para\'iba, Caixa Postal 5008, 58051-900, Jo\~ao Pessoa-PB, Brazil.}

\author{K. Bakke}
\email{kbakke@fisica.ufpb.br}
\affiliation{Departamento de F\'isica, Universidade Federal da Para\'iba, Caixa Postal 5008, 58051-900, Jo\~ao Pessoa-PB, Brazil.}

\begin{abstract}
We investigate noninertial effects on the scattering problem of a nonrelativistic particle in the cosmic string spacetime. By considering the nonrelativistic limit of the Dirac equation we are able to show, in the regime of small rotational frequencies, that the phase shift has two contribution: one related to the noninertial reference frame, and the other, due to the cosmic string conical topology. We also show that both the incident wave and the scattering amplitude are altered as a consequence of the noninertial reference frame and depend on the rotational frequency.
\end{abstract}

\keywords{topological scattering, Noninertial effects, topological defect, cosmic string}
\pacs{}

\maketitle

\section{Introduction}

Effects of rotation on nonrelativistic classical systems have been reported in the literature \cite{mechanics,statistical} and inspired works in the context of quantum mechanics \cite{Mashhoon:1988zz,rizzi2013relativity, 2003quant.ph..5081A, sagnac1, sagnac2,Anandan:1977ra,Iyer:1982ah,POST:1967zz, Sakurai:1980te}. The most known effects that stem from rotation are associated with the appearance of geometric quantum phases such as the Sagnac effect \cite{sagnac1, sagnac2,POST:1967zz}, the Mashhoon effect \cite{Mashhoon:1988zz} and the Aharonov-Carmi geometric phase \cite{Aharonov1973}. Another effect has been pointed out in Refs. \cite{Page:1975zz, Werner:1979gi, Hehl:1990nf}, where a contribution to the energy levels stems from the coupling between the angular momentum and the angular velocity of the rotating frame. Based on Refs. \cite{mechanics,statistical}, it has been investigated in Ref. \cite{2015PhLA..379...11D} effects of rotation on a nonrelativistic quantum system by writing the Hamiltonian operator in the form: $\hat{H}=\hat{H}_{0}-\vec{\Omega}\cdot\hat{L}$, where $\hat{H}_{0}$ corresponds to the Hamiltonian operator in the absence of rotation, $\vec{\Omega}$ is the angular velocity of the rotating frame and $\hat{L}$ is the angular momentum operator. Recently, this proposal has been explored in neutral particles systems with a magnetic quadrupole moment \cite{Fonseca:2016iod}. Other contexts that have investigated rotating effects are the quantum Hall effect \cite{fisher}, spintronics \cite{PhysRevLett.106.076601,Chowdhury2013358, PhysRevB.84.104410} and Bose-Einstein condensation \cite{PhysRevA.76.023410}.

On the other hand, in relativistic classical systems, Landau and Lifshitz \cite{CTF} showed that the line element of the Minkowski spacetime becomes singular at large distances for a system in a uniformly rotating frame. This particular aspect has attracted interests in the quantum field theory with studies in Dirac fields \cite{Iyer:1982ah,Soares:1995cj}, Casimir effect \cite{Mota:2014nka}, scalar bosons \cite{Castro:2015zca}, DKP equation \cite{Hassanabadi:2015rla}, electroweak interactions \cite{Dvornikov:2015dna}, neutrino interactions \cite{Dvornikov:2015tza} and the Dirac oscillator \cite{Bakke:2012zz}. This geometrical approach proposed by Landau and Lifshitz \cite{statistical} has also been explored in the nonrelativistic limit of the Dirac equation with the confinement of a neutral particle to a hard-wall confining potential \cite{Bakke2013} and in the presence of torsion \cite{Bakke:2014aza}.

The aim of this work is to investigate effects of rotation on the nonrelativistic quantum scattering of a spin-$1/2$ particle in the cosmic string spacetime. It is well-known that cosmic strings are line-like topological defects predicted in the context of some gauge extensions of the Standard Model of particle physics, and give rise to a variety of cosmological, astrophysical and gravitational phenomena \cite{VS,hindmarsh,Copeland:2011dx}. For instance, a cosmic string can emit gravitational waves and high energy cosmic rays \cite{Mota:2014uka} that makes possible the product of the cosmic string linear energy density ($\mu$) by the Newton's gravitational constant ($G$) to be constrained. In special, an idealized structureless and very long straight cosmic string produces a topologically conical spacetime with an angle deficit, on the plane perpendicular to it, given by $\Delta\varphi = 8\pi G\mu$ \cite{VS}. Besides, the axial symmetry of the cosmic string spacetime\- provides an interesting way of studying scattering problems since it can be viewed as a point-like particle with mass $\mu = M$ in $(2+1)$-dimensions. In fact, it has been widely shown that the interaction between two-point particles can be described in terms of the dynamics of a particle on a $(2+1)$-conical spacetime \cite{'tHooft:1988yr}. In this scenario, the scattering of the point particle is caused by the nontrivial topology of the conical spacetime. With this spirit, the nonrelativistic and relativistic scattering problem of a particle in the $(2+1)$-dimensional cosmic string spacetime were considered in Refs. \cite{Deser:1988qn,Gerbert}, where the authors showed that the scattering amplitude depends crucially on the cosmic string parameter $\alpha=1-4G\mu$. Furthermore, extensions of these pioneers works have been realized in Ref. \cite{Mota:2016eoi}, where it is analyzed the scattering problem of a relativistic bosonic particle under a nontrivial boundary condition. In Ref. \cite{Spinally:2000ii} is considered the high delta-like curvature contribution for the scattering\- amplitude and the nonrelativistic time-dependent generalization has been performed in Ref. \cite{Alvarez:1995fs}.

The structure of this paper is as follows: in section II, we begin by making a brief review of a Dirac particle in the Fermi-Walker reference frame, and thus we show the nonrelativistic limit of the Dirac equation; in section III, we discuss the nonrelativistic quantum scattering of a spin-$1/2$ particle in the presence of noninertial effects and a topological defect; in section IV, we present our conclusions.

\section{Nonrelativistic limit of the Dirac equation in the Fermi-Walker referential frame revisited}

In this section, we start by making a brief review of a Dirac particle in the cosmic string spacetime under the effects of rotation and, in the following, we discuss the nonrelativistic limit of the Dirac equation. The cosmic string spacetime is a topological defect spacetime described by the line element:
\begin{eqnarray}
ds^{2} =-dt^{2}+dr^{2}+\alpha^{2}r^{2}d\varphi^{2}+dz^{2},
\label{2.1}
\end{eqnarray}
where $\alpha=1-4G\mu$ is the parameter associated with the deficit of angle and $0\leq\varphi<2\pi$ \cite{Kibble:1976sj,Katanaev:1992kh,Katanaev:1998xw,Starobinskii}. Next, by following Ref. \cite{statistical}, we can make a coordinate transformation given by $\varphi\rightarrow\varphi+\varpi\,t$, where $\varpi$ corresponds to the constant angular velocity of the rotating frame that rotates around the $z$-axis, then, the line element of the cosmic string spacetime (\ref{2.1}) becomes \cite{Bakke:2012zz}: 
\begin{eqnarray}
ds^{2}=-\left(1-\varpi^2\alpha^{2}r^{2}\right)\,dt^{2}+2\varpi\alpha^{2}r^{2}d\varphi\,dt+dr^{2}+\alpha^{2}r^{2}d\varphi^{2}+dz^{2}.
\label{FWM}
\end{eqnarray}

As pointed out in Refs. \cite{Bakke2013,Bakke:2012zz}, the line element of the cosmic string under the effects of rotation becomes defined in the range 
\begin{eqnarray}
0\,<\,r\,<\,\frac{1}{\alpha\,\varpi},
\label{2.2}
\end{eqnarray}
which means that there is a restriction on the values of the radial coordinate, otherwise, for $r\geq\frac{1}{\alpha\,\varpi}$, the line element (\ref{FWM}) becomes positive, which is not admissible \cite{statistical}. As discussed in Ref. \cite{Bakke:2012zz}, if $r\geq\frac{1}{\alpha\,\varpi}$ we would have a particle with velocity greater than the velocity of light, which agrees with Ref. \cite{statistical} in the Minkowski spacetime, since we can recover the line element of the Minkowski spacetime by taking $\alpha\rightarrow1$. In recent years, the restriction (\ref{2.2}) imposed by the rotating effects has been explored in the context of quantum mechanics, where it has been used to impose that the wave function of the Dirac particle vanishes at $r\rightarrow1/\alpha\,\varpi$. As examples, the Dirac oscillator \cite{Bakke:2012zz} and the Casimir effect \cite{Mota:2014nka}.

In order to work with the Dirac equation, we need to build local reference frame of the observers is through a noncoordinate basis $\hat{\theta}^{a}=e^{a}_{\,\,\,\mu}\left(x\right)\,dx^{\mu}$, where the components $e^{a}_{\,\,\,\mu}\left(x\right)$ are the well-known tetrads that satisfy the relation $g_{\mu\nu}\left(x\right)=e^{a}_{\,\,\,\mu}\left(x\right)\,e^{b}_{\,\,\,\nu}\left(x\right)\,\eta_{ab}$, with $\eta_{ab}=\mathrm{diag}\left(-\,+\,+\,+\right)$ as being the Minkowski metric tensor \cite{birrell1984quantum,nakahara2003geometry}. Then, let us take a Fermi-Walker reference frame \cite{thorne2000gravitation,weinberg1972gravitation,Bakke:2012zz}:
\begin{eqnarray}
\hat{\theta}^{0}=dt;\,\,\,\,\hat{\theta}^{1}=dr;\,\,\,\,\hat{\theta}^{2}=\varpi\,\alpha\,r\,dt+\alpha\,r\,d\varphi;\,\,\,\hat{\theta}^{3}=dz.
\label{2.3}
\end{eqnarray}

Thereby, the Dirac equation in the Fermi-Walker reference frame (\ref{2.3}) is written in form:
\begin{eqnarray}
i\gamma^{\mu}\,\partial_{\mu}\Psi+i\gamma^{\mu}\,\Gamma_{\mu}\left(x\right)\Psi=m\Psi,
\label{2.4}
\end{eqnarray}
where $\Gamma_{\mu}\left(x\right)=\frac{i}{4}\,\omega_{\mu ab}\left(x\right)\,\Sigma^{ab}$ is the spinorial connection \cite{birrell1984quantum,nakahara2003geometry}, with $\omega_{\mu ab}\left(x\right)$ as being the connection 1-form or spin connection, whose components can be obtained by solving the Maurer-Cartan structure equations in the absence of torsion $d\hat{\theta}^{a}+\omega^{\,\,\,a}_{\mu\,\,\,b}\left(x\right)\,dx^{\mu}\wedge\,\hat{\theta}^{b}=0$ \cite{nakahara2003geometry}. Furthermore, the term $\Sigma^{ab}=\frac{i}{2}\left[\gamma^{a},\,\gamma^{b}\right]$ is determined by the standard Dirac matrices in Minkowski spacetime \cite{greiner1990relativistic}: 
\begin{eqnarray}
\gamma^{0}= 
\begin{pmatrix} 
1  &  0 \\ 
0  &  -1  
\end{pmatrix};\;\;\;\;\;\;\;\;\;\;
\gamma^{i}=
\begin{pmatrix} 
0  &  \sigma^{i} \\ 
-\sigma^{i}  &  0
\end{pmatrix}, 
\label{2.5}
\end{eqnarray}
where $\sigma^{i}$ corresponds to the Pauli matrices. It is easy to check that the second term on the left-hand side of the Dirac equation (\ref{2.4}) becomes $i\gamma^{\mu}\,\Gamma_{\mu}\left(x\right)=i\frac{\gamma^{1}}{2r}$. Then, by following Refs. \cite{Bakke2013,greiner1990relativistic}, we can write the solution to the Dirac equation (\ref{2.4}) as $\Psi=e^{-i\,m\,t}\,\left(\phi\,\,\,\chi\right)^{T}$, where $\phi$ is considered to be the ``large'' component and $\chi$ is considered to be the ``small'' component, and thus we can take the nonrelativistic limit of the Dirac equation. In this way, after some calculations, we obtain the Schr\"odinger equation:
\begin{eqnarray}
i\frac{\partial\phi}{\partial t}=-\frac{1}{2m}\left[\frac{\partial^{2}}{\partial r^{2}}+\frac{1}{r}\frac{\partial}{\partial r}+\frac{1}{\alpha^{2}\,r^{2}}\frac{\partial^{2}}{\partial\varphi^{2}}+\frac{\partial^{2}}{\partial z^{2}}\right]\phi+\frac{i}{2m}\frac{\sigma^3}{\alpha\,r^{2}}\frac{\partial\phi}{\partial\varphi}+\frac{1}{8m\,r^{2}}\,\phi+i\varpi\frac{\partial\phi}{\partial\varphi}. 
\label{RDE}
\end{eqnarray}
Note that, by taking $\alpha\rightarrow1$, we recover the Schr\"odinger equation in the Fermi-Walker reference frame obtained Ref. \cite{Bakke2013} in the absence of a topological defect. A particular solution to Eq. (\ref{RDE}) is obtained by observing that $\phi$ is an eigenfunction of $\sigma^{3}$ ($\sigma^{3}\phi=\pm\phi=s\phi$, where $s=\pm1$) and the operators $\hat{J}_{z}=-i\partial_{\varphi}$ \cite{Schluter:1983ep} and $\hat{p}_{z}=-i\partial_{z}$ commute with the Hamiltonian operator defined on the right-hand side of Eq. (\ref{RDE}). Hence, we have
\begin{eqnarray}
\phi\left(t,\,r,\,\varphi,\,z\right)=e^{-i\mathcal{E}t}e^{i\left(l+\frac{1}{2}\right)\varphi}\,e^{ip_{z}\,z}\,R_{l,\,s}\left(r\right).
\label{GS}
\end{eqnarray}
By substituting the wave function (\ref{GS}) into Eq. (\ref{RDE}), we obtain the following radial equation:
\begin{eqnarray}
\frac{\partial^{2}R_{l,\,s}}{\partial r^{2}} + \frac{1}{r}\frac{\partial R_{l,\,s}}{\partial r}+\left(\eta_{l}^{2}-\frac{\nu_{s,q}^{2}}{\alpha^{2}\,r^{2}}\right)R_{l,\,s}=0.
\label{RE}
\end{eqnarray}
where we have taken $p_{z}=0$ and defined the parameters
\begin{eqnarray}
\eta_{l}^{2}&=&\eta^2 + 2m\varpi\left(l+\frac{1}{2}\right);\nonumber\\
[-2mm]\label{2.6}\\[-2mm]
\nu_{s,q}&=&l+\frac{1}{2}-\frac{s}{2q}=l+\beta_{q}\nonumber,
\end{eqnarray}
where $\eta^2 = 2m\mathcal{E}$, $q=1/\alpha$ and $\beta_{q}=\left(q-s\right)/2q$. Note that the second order differential equation (\ref{RE}) is known in the literature as the Bessel differential equation \cite{arfken2011mathematical,abramowitz1966handbook}. The general solution to Eq. (\ref{RE}) is given by $R_{l,\,s}\left(r\right)=A\,J_{\nu_{s,q}}\left(\eta_{l}\,r\right)+B\,N_{\nu_{s,q}}\left(\eta_{l}\,r\right)$, where $J_{\nu_{s,q}}\left(\eta_{l}\,r\right)$ and $N_{\nu_{s,q}}\left(\eta_{l}\,r\right)$ are the Bessel functions of the first kind and second kind, respectively. Since we wish to deal with a regular solution at the origin, then, we consider $B=0$ and write the solution to Eq. (\ref{RE}) in the form:
\begin{eqnarray}
R_{l,\,s}\left(r\right)=A\,J_{q\,\left|\nu_{s,q}\right|}\left(\eta_{l}\,r\right).
\label{RS}
\end{eqnarray}

Now, we are able to investigate the effects of rotation on the nonrelativistic quantum scattering of a spin-$1/2$ particle in the presence of a topological defect.

\section{Topological scattering in the Fermi-Walker reference frame}

\subsection{The partial wave expansion approach}

Since the cosmic string spacetime has an axial symmetry, to perform our analysis, we will omit the $z$-component of the solution (\ref{GS}) and only consider the azimuthal and radial parts. Bearing that in mind, the scattering solution of the Dirac equation is constructed such that there is a nonzero phase shift $\delta_{l}$ \cite{Deser:1988qn,Gerbert}, i.e,
\begin{equation}
\phi(r,\varphi) = \sum_{l=-\infty}^{\infty}e^{i\left(\delta_{l} + \frac{|\nu_{s,1}|\pi}{2}\right)}R_{l,s}(r)e^{i\left(l + \frac{1}{2}\right)\varphi},
\label{exem}
\end{equation}
where $\nu_{s,1}$ is given by Eq. (\ref{2.6}), with $q=1$. This solution correctly provides the large-distance scattering behaviour
\begin{equation}
\phi(r,\varphi)\xrightarrow[r \to \infty] {} e^{i\eta r\cos\varphi}+\sqrt{\frac{i}{r}}f(\varphi)e^{i\eta r},
 \label{AE}
\end{equation}
where
\begin{equation}
f(\varphi)=\frac{1}{\sqrt{-2\pi\eta}}\sum_l\left(e^{2i\delta_l}-1\right)e^{i\left(l + \frac{1}{2}\right)\varphi},
\label{SA}
\end{equation}
is the scattering amplitude. Note that Eqs. (\ref{exem})-(\ref{SA}) represent the partial wave expansion approach for scattering problems \cite{Deser:1988qn,Gerbert}.

If there is really scattered waves, asymptotically,  the radial solution in Eq. (\ref{RS}) will differ from that of Minkowski spacetime by a phase shift $\delta_l$, that is,
\begin{equation}
 R_{l,s}(r)\xrightarrow[r \to \infty] {} \sqrt{\frac{2}{\pi\eta r}} \cos\left(\eta r-\frac{q|\nu_{s,1}|\pi}{2}-\frac{\pi}{4}+\delta_{l}\right).
\label{WFAS}
\end{equation}

Thereby, by comparing the asymptotic behaviour of Eq. (\ref{RS}), when $q=1$ and $\varpi=0$, with the expression above we see that
\begin{equation}
 \delta_{l} = \delta\eta_l + \delta_q,
\label{PS}
\end{equation}
where 
\begin{eqnarray}
\delta\eta_l &=& \eta_l - \eta,\nonumber\\
&=& \sqrt{2m\left[\mathcal{E} + \varpi\left(l + \frac{1}{2}\right)\right]} - \eta,
\label{eta}
\end{eqnarray}
is the phase shift associated with the existence of a nonzero rotational frequency $\varpi$, characterizing the noninertial reference frame. As we are working in the domain of the nonrelativistic limit of the Dirac equation (\ref{2.4}) it is reasonable to consider only small frequencies, that is, $\mathcal{E}\gg\varpi$. This assumption provides us with the following approximation for Eq. (\ref{eta}):
\begin{eqnarray}
\delta\eta_l \approx \omega_{\rm{eff}}\left(l + \frac{1}{2}\right), \,\,\,\,\,\,\,\,\,\, \omega_{\rm{eff}}=\sqrt{\frac{m}{2\mathcal{E}}}\varpi.
\label{eta2}
\end{eqnarray}
In addition, the second term on the right hand side of Eq. (\ref{PS}) is the phase shift associated with the cosmic string topology given by
\begin{equation}
\delta_l=\frac{\pi}{2}(|\nu_{s,1}|-q|\nu_{s,q}|)=\left\{ \begin{array}{l}-\frac{1}{2}\left(l + \frac{1}{2}\right)\omega_{\rm cs},\,\,\,\,\,\mathrm{for}\,\,\, \,l\ge 0\,,\\
\,\,\,\,\,\frac{1}{2}\left(l + \frac{1}{2}\right)\omega_{\rm cs},\,\,\,\,\,\mathrm{for}\,\,\,\, l< 0\,,
\end{array}\right.
\label{CSPS}
\end{equation}
where $\omega_{\rm cs} = \pi(q -1)$ is the classical scattering. One should note that for the case $\varpi=0$, the phase shift (\ref{PS}) is only due to the presence of the cosmic string. This case was firstly investigated in \cite{Gerbert}, where the authors obtained Eq. (\ref{CSPS}).

The natural step now would be to use the phase shift (\ref{PS}) along with Eq. (\ref{SA}) to calculate the scattering amplitude. However, as already showed in Refs. \cite{Deser:1988qn,Gerbert}, Eq. (\ref{SA}) leads to the wrong scattering amplitude, given in terms of delta functions. By using the Schl$\ddot{\mathrm{a}}$fli representation for the Bessel function $J_\nu(z)$, the authors in \cite{Deser:1988qn,Gerbert} obtained, using solution (\ref{exem}), the correct separation that provides the correct and finite scattering amplitude, although they did that using a rather complicated contour integration. Following the same approach adopted in \cite{Mota:2016eoi}, we wish to obtain a summation formula for Eq. (\ref{exem}) which will provide the correct scattering amplitude in an easier and more elegant way.

\subsection{The scattering problem}

Let us perform the Wick-like rotation, $\eta_l r = iz_l$, in the radial solution (\ref{RS}). Upon using the well-known relation $J_n(iz) = e^{i\frac{n\pi}{2}} I_n(z)$ and the phase shift (\ref{PS}), Eq. (\ref{exem}) becomes
\begin{equation}
\phi(r,\varphi) = e^{i\frac{\varphi_s\pi}{2}}S(z_l),
\label{IR}
\end{equation}
where $\varphi_s = \varphi + (1 - s)$ and 
\begin{equation}
S(z_l) = \sum_{l=-\infty}^{\infty}e^{ir\delta\eta_l}I_{q(l + \beta_q)}(z_l)e^{il\vartheta},
\label{sum}
\end{equation}
with $\vartheta = \varphi + \pi$. Since we are considering the case $q>1$ (cosmic string), $\beta_q$ is defined in the interval $0<\beta_q<1$ . As a consequence, we could safely turn the absolute value $|(l + \beta_q)|$ into only $(l + \beta_q)$ \cite{Braganca:2014qma}.

The sum in Eq. (\ref{sum}) can be evaluated by using the integral representation for the Bessel function $I_n(z)$ \cite{abramowitz1966handbook}, i.e,
\begin{equation}
 I_{\epsilon_l}(z_l) = \frac{1}{\pi}\int_0^{\pi}dy\cos(\epsilon_ly)e^{z_l\cos y} - \frac{\sin(\pi\epsilon_l)}{\pi}\int_0^{\infty}dye^{-z_l\cosh y - \epsilon_ly},
 \label{BIR}
\end{equation}
where $\epsilon_l = q(l + \beta_q)$. By noting that $z_l = -ir(\eta + \delta\eta_l)$, we get from Eqs. (\ref{sum}) and (\ref{BIR}) 
\begin{equation}
S(z) =S_{\rm in}(r,\vartheta) \label{sum2} + S_{\rm sc}(r,\vartheta),
\end{equation}
where
\begin{eqnarray}
S_{\rm in}(r,\vartheta)=\frac{1}{\pi}\int_0^{\pi}dye^{z\cos y}\sum_{l=-\infty}^{\infty}\cos(\epsilon_l y)e^{il\vartheta}e^{2ir\delta\eta_l\sin^2\left(\frac{y}{2}\right)},
\label{IW}
\end{eqnarray}
is the contribution that provides the incident wave, as we will see below, and 
\begin{eqnarray}
S_{\rm sc}(r,\vartheta)=-\frac{1}{\pi}\int_0^{\infty}dye^{-z\cosh(y)}\sum_{l=-\infty}^{\infty}e^{il\vartheta - \epsilon_ly}\sin(\pi\epsilon_l)e^{2ir\delta\eta_l\cosh^2\left(\frac{y}{2}\right)},
\label{SW}
\end{eqnarray}
is the contribution that provides the scattered wave, with $z = -i\eta r$.

It is already known that when $\eta r\rightarrow\infty$, the main contribution for the integrals in Eqs. (\ref{IW}) and (\ref{SW}) comes from small values of $y$ \cite{Deser:1988qn,Gerbert,Mota:2016eoi}. Thereby, using Eq. (\ref{eta2}), as a first approximation for Eq. (\ref{IW}), one has
\begin{eqnarray}
S_{\rm in}(r,\vartheta)\approx\frac{1}{\pi}\int_0^{\pi}dye^{-z\frac{\omega_{\rm{eff}}}{\eta}\sin^2\left(\frac{y}{2}\right)}e^{z\cos y}\sum_{l=-\infty}^{\infty}\cos(\epsilon_l y)e^{il\vartheta},
\label{IW2}
\end{eqnarray}
where $e^{-2z\frac{\omega_{\rm{eff}}}{\eta}l\sin^2\left(\frac{y}{2}\right)}\approx 1$, since $\frac{\omega_{\rm{eff}}}{\eta}, y\ll 1$. The summation and integral above can be calculated in a similar way as in Ref. \cite{deMello:2014ksa,Mota:2016eoi}, providing
\begin{eqnarray}
S_{\rm in}(r,\vartheta)=\frac{e^{i\frac{r\omega_{\rm{eff}}}{2}}}{q}\sum_ne^{-ir\eta\left(1 + \frac{\omega_{\rm{eff}}}{2\eta}\right)\cos\left[\frac{(\vartheta-2\pi n)}{q}\right]}e^{-i\beta_q(\vartheta-2\pi n)},
\label{IW3}
\end{eqnarray}
which is the incident wave. One can see that when $\varpi=0$ $(\omega_{\rm{eff}}=0)$ we recover the incident wave obtained by the authors in \cite{Gerbert}, that is, the incident wave due to the scatting problem in the cosmic string spacetime.

Let us now consider an approximated expression for Eq. (\ref{SW}). Once again, using Eq. (\ref{eta2}), one obtains 
\begin{eqnarray}
S_{\rm sc}(r,\vartheta) &=& -\frac{e^{z \frac{\omega_{\rm{eff}}}{2\eta}}}{\pi}\int_0^{\infty}dye^{-z\left(1 + \frac{\omega_{\rm{eff}}}{2\eta}\right)\cosh(y)}\sum_{l=-\infty}^{\infty}e^{il\Delta\vartheta - \epsilon_ly}\sin(\pi\epsilon_l),\nonumber\\
&\approx& -\frac{e^{-z}}{\pi}\int_0^{\infty}dye^{-z\left(1 + \frac{\omega_{\rm{eff}}}{2\eta}\right)\frac{y^2}{2}}\sum_{l=-\infty}^{\infty}e^{il\Delta\vartheta - \epsilon_ly}\sin(\pi\epsilon_l),
\label{SW5}
\end{eqnarray}
where 
\begin{eqnarray}
\Delta\vartheta &=&\vartheta + 2r\omega_{\rm{eff}}\cosh^2(y/2),\nonumber\\
&\approx&\vartheta + 2r\omega_{\rm{eff}}.
\label{angle}
\end{eqnarray}

The summation in Eq.  (\ref{SW5}) can be performed using the same procedure as in \cite{deMello:2014ksa,Mota:2016eoi}. This provides 
\begin{eqnarray}
S_{\rm sc}(r,\vartheta)=-\frac{e^{-z}}{2\pi i}\sum_{j=+,-}je^{j\pi iq\beta_q}\int_{0}^{\infty}dy\frac{\cosh[qy(1-\beta_q)]-\cosh(q\beta_q y)e^{-i\left(\Delta\vartheta+jq\pi\right)}}{e^{z\left(1 + \frac{\omega_{\rm{eff}}}{2\eta}\right)\frac{y^2}{2}}[\cosh(qy)-\cos\left(\Delta\vartheta+jq\pi\right)]}.
\label{SW2}
\end{eqnarray}
Note that the integral in $y$ gives the correct large-distance behaviour to get the scattering amplitude. In order to see that Eq. (\ref{SW2}) really provides the correct asymptotic scattering behaviour one needs to expand the $y$-dependent functions and consider only the leading term. By doing so one obtains
\begin{eqnarray}
S_{\rm sc}(r,\Delta\vartheta)&=&-\frac{1}{2\pi i}\sum_{j=+,-}je^{j\pi iq\beta_q}\frac{1-e^{-i\left(\Delta\vartheta+jq\pi\right)}}{[1-\cos\left(\Delta\vartheta+jq\pi\right)]}e^{-z}\int_0^{\infty}e^{-z\left(1 + \frac{\omega_{\rm{eff}}}{2\eta}\right)\frac{y^2}{2}}dy,\nonumber\\
&=&\sqrt{\frac{i}{r}}f(\Delta\vartheta)e^{i\eta r},
\label{SW3}
\end{eqnarray}
where
\begin{eqnarray}
f(\Delta\vartheta)&=&\frac{1}{\sqrt{2\pi\eta}}\frac{1}{\cos\omega_{\rm cs} +\cos\Delta\vartheta}\left(\sin\bar{\omega}-\sin(q\pi\beta_q)e^{-i\Delta\vartheta}\right)\left(1 + \frac{\omega_{\rm{eff}}}{2\eta}\right)^{-\frac{1}{2}},\nonumber\\
&=&\frac{2i\cos\left(\frac{S}{2}\right)}{\sqrt{2\pi\eta}}\frac{\sin\left(\frac{\omega_{\rm cs} }{2}\right)\sin\left(\frac{\vartheta + 2r\omega_{\rm{eff}}}{2}\right)e^{-i\frac{(\vartheta + 2r\omega_{\rm{eff}})}{2}}}{\cos\omega_{\rm cs} +\cos(\vartheta + 2r\omega_{\rm{eff}})}\left(1 + \frac{\omega_{\rm{eff}}}{2\eta}\right)^{-\frac{1}{2}},
\label{SA5}
\end{eqnarray}
is the correct scattering amplitude for $\Delta\vartheta\neq q\pi$, with $\bar{\omega}=\omega_{\rm cs} - q\pi\beta_q$, $q\pi\beta_q = \frac{\omega_{\rm cs}}{2} +  \frac{S}{2}$ and $S=\pi (1-s)$. Within the approximation we have taken for Eqs. (\ref{IW2}) and (\ref{SW5}), we can clearly see the dependency of Eq. (\ref{SA5}) with the rotational frequency $\varpi $, or equivalently with $\omega_{\rm{eff}}$. This dependency only appears if we also consider the nontrivial topology of the cosmic string spacetime, otherwise, by making $q=1$ in Eq. (\ref{SA5}) there will be no scattered wave and therefore no effect of the noninertial reference frame will be evident.

In the limit $\varpi =0$ we recover the result \footnote{Eq. (\ref{SA5}), for $\varpi =0$, is an alternative way of writing Eq. (5.25) of \cite{Gerbert}.} of Ref. \cite{Gerbert}. However, we point out that while in \cite{Gerbert} the authors obtained the same scattering amplitude for $s=\pm 1$ here we can see that they are the same up to a minus sign, which is a consequence of whether $S=0$ or $S=2\pi$. The reason for this is that we use the more correct approach to calculate the scattering amplitude, in contrast with the approach adopted in \cite{Gerbert} where they made use of Eq. (\ref{SA}). In fact, as pointed out by the author in Ref. \cite{Hagen:1989va} one should find a way of first solving the summation in (\ref{exem}) and then take the large-distance limit to analise the scattering problem. Thus, by doing so, we do not get for instance the delta's contributions in Eq. (\ref{SA5}) as in \cite{Gerbert}.

One should note that the scattering amplitude has an apparent divergency at $\Delta\vartheta=\pm q\pi$. Nevertheless, we can show, in a similar way as it was done in \cite{Mota:2016eoi}, that this is not the case. To obtain the correct finite expression for the scattering amplitude at these values, one needs to reconsider Eq.  (\ref{SW2}) and expand the cosh's up to O($y^2$). By doing that one gets
\begin{eqnarray}
S_{\rm sc}(r,\pm q\pi)\xrightarrow[\eta r \to \infty] {}&-&\frac{1}{2\pi i}\left[\mp(1-2\beta_q)e^{\mp i\pi q\beta_q}+i\frac{e^{\pm i\pi q\beta_q}e^{\mp i\omega_{\rm cs}}}{\sin\omega_{\rm cs}}\right]e^{-z}\int_0^{\infty}e^{-z\left(1 + \frac{\omega_{\rm{eff}}}{2\eta}\right)\frac{y^2}{2}}dy,\nonumber\\
&=&\sqrt{\frac{i}{r}}\bar{f}(\pm q\pi)e^{i\eta r},
\label{E13}
\end{eqnarray}
where
\begin{eqnarray}
f(\Delta\vartheta=\pm q\pi)&=&\frac{1}{\sqrt{2\pi\eta}}\left(-\frac{e^{\pm i\pi q\beta_q}}{2}\cot\omega_{\rm cs}-\sin(\pi q\beta_q)\pm i\beta_q e^{\mp i\pi q\beta_q}\right)\left(1 + \frac{\omega_{\rm{eff}}}{2\eta}\right)^{-\frac{1}{2}},\nonumber\\
&=&\frac{e^{i\frac{S}{2}}}{\sqrt{2\pi\eta}}\left(-\frac{e^{\pm i\frac{\omega_{\rm cs}}{2}}}{2}\cot\omega_{\rm cs}-\sin\left(\frac{\omega_{\rm cs}}{2}\right)\pm i\beta_q e^{\mp i\frac{\omega_{\rm cs}}{2}}\right)\left(1 + \frac{\omega_{\rm{eff}}}{2\eta}\right)^{-\frac{1}{2}},
\label{E14}
\end{eqnarray}
which is the finite scattering amplitude valid for $q\neq 1$ and calculated at $\Delta\vartheta=\pm q\pi$. The latter provides an interesting result for the rotational frequency, that is, 
\begin{eqnarray}
\varpi = \sqrt{\frac{2\mathcal{E}}{m}}\frac{(\omega_{\rm cs} - \varphi)}{2r},
\label{fre}
\end{eqnarray}
for a fixed energy $\mathcal{E}$ and defined in the interval $0< r < \frac{q}{\varpi}$. This expression for the rotational frequency taken at $\varphi =  \omega_{\rm cs}$ is zero while at $\varphi =  - \omega_{\rm cs}$ is given by $\varpi = \sqrt{\frac{2E}{m}}\frac{\omega_{\rm cs}}{r}$. In other words, the rotational frequency $\varpi$ is directly related to the topology of the cosmic string spacetime through $\omega_{\rm cs}$.

Let us now point out another interesting feature of Eq. (\ref{E14}). It will provide a somewhat different expression for the scattering amplitude, depending on the value of the spin $s=\pm 1$. That is, because of the last term on the right hand side of Eq. (\ref{E14}), the expressions for $s=\pm 1$ will differ for more than a minus sign, different from what happens in Eq. (\ref{SA5}). Moreover, to the best of our knowledge, the scattering amplitude (\ref{E14}) calculated at $\Delta\vartheta=\pm q\pi$ had not been obtained so far for the case $\varpi=0$. The authors in \cite{Gerbert} only obtained the expression (\ref{SA5}) for $\Delta\vartheta\neq\pm q\pi$. 

It is worth noting that although we have considered here the nonrelativistic limit of the Dirac equation, the main results for the scattering problem such as phase shift, scattering amplitude and the incident wave, in the case $\varpi = 0$, are the same as the ones obtained in \cite{Gerbert} in the relativistic regime. The difference appears only in the expression for the energy, which in the nonrelativistic case is given by Eq. (\ref{2.6}).

\section{conclusions}

We have considered noninertial effects on the scattering problem of a spinorial particle on a $(2 + 1)$-dimensional cosmic string spacetime. Upon taking the nonrelativistic limit (\ref{RDE}) of the Dirac equation (\ref{2.4}), we have been able to  find its general solutions (\ref{GS}) and the corresponding energy levels of the system which depend on the rotational frequency, $\varpi$, of the rotational reference frame. The nonrelativistic limit have allowed us to study the more tractable case of small frequencies.

In the context of the partial wave expansion approach for the scattering problem we have also obtained the phase shift (\ref{PS}) and shown it has two contributions, one corresponding to the noninertial effects contribution (\ref{eta}) and another one corresponding to the conical topology of the cosmic string spacetime (\ref{CSPS}). The latter was previously obtained in \cite{Gerbert} and we emphasise that although the authors obtained this phase shift in the relativistic regime, for $\varpi=0$, all the results for the scattering problem, i.e, phase shift, scattering amplitude etc., are the same in the nonrelativistic case, except for the fact that the expression for the energy assume the nonrelativistic form (\ref{2.6}). 

Using the small frequencies approximation in Eq. (\ref{eta2}) we have also been able to obtain both the incident wave (\ref{IW3}) and the scattered wave (\ref{SW3}) by adopting the integral representation (\ref{BIR}) of the Bessel function $I_n(x)$. We have shown that the incident wave and the scattering amplitude (\ref{SA5}) depend on $\varpi$, through the parameter $\omega_{\rm{eff}}$, defined in Eq. (\ref{eta2}). We point out that Eq. (\ref{SA5}) provides, for $s=\pm 1$, the same expressions up to a minus sign, in contrast to the result of \cite{Gerbert} where the authors showed the expressions are exactly the same. This is due to the fact we have first performed the summation in $l$ in Eq. (\ref{exem}), using a summation formula previously obtained in \cite{deMello:2014ksa}, and only afterwards taken the large-distance limit. This has been pointed out by the author in \cite{Hagen:1989va} to be the more correct way of tackling the problem.

The apparent divergency at $\Delta\vartheta = \pm q\pi$ present in Eq. (\ref{SA5}) was shown to be finite in Eq. (\ref{E14}) after we expanded the cosh's in Eq. (\ref{SW2}) up to $O(y^2)$. The scattering amplitude (\ref{E14}) at $\Delta\vartheta = \pm q\pi$ was not obtained by Ref. \cite{Gerbert}  and, as far as we know, is a new result derived here. Because of the last term on the right hand side of Eq. (\ref{E14}) , the expression for $s=\pm 1$ will not differ only by a minus sign. Moreover, as a consequence of Eq. (\ref{E14}) taken at $\Delta\vartheta = \pm q\pi$ we also derived, using Eq. (\ref{angle}) , an interesting expression for the rotational frequency given by Eq. (\ref{fre}) .

It is also possible to conduct an investigation of noninertial effects on the scattering of a bosonic particle propagating in $(2 + 1)$-dimensional cosmic string spacetime. In this case, similarly to Eqs. \eqref{SA5} and \eqref{E14}, we expect to obtain a dependence of the scattering amplitude on the rotational frequency, as well as an analogous formula to Eq.  (\ref{fre}). An additional possibility is to consider noninertial effects in another geometry, like the one that describes the spacetime with a space-like dislocation \cite{valdir,Bakke:2014aza,vb}, a linear topological defect associated with torsion which is also found in the context of theories of solid and crystal continuum media \cite{Puntigam:1996vy,Letelier:1995ze}. In this context, the scattering problem of both fermionic and bosonic particles can be analysed.

\acknowledgments{We would like to thank CNPq (Conselho Nacional de Desenvolvimento Cient\'ifico e Tecnol\'ogico - Brazil) for financial support.}

\end{document}